\documentclass[11pt]{article}
\usepackage[a4paper,margin=1in]{geometry}

\usepackage{amsmath,amssymb,bm}
\usepackage{algorithm}
\usepackage{algorithmic}
\usepackage{graphicx}
\usepackage{hyperref}
\usepackage{subcaption}
\usepackage{tikz}
\usepackage{xcolor}
\usetikzlibrary{arrows.meta,positioning,fit,shapes,calc}
\usepackage{booktabs}
\usepackage{enumitem}
\usepackage{placeins}
\usepackage{float}

\hypersetup{
    colorlinks=true,
    linkcolor=black,
    citecolor=black,
    urlcolor=blue
}

\newcommand{\Groupname}{\textsc{Salient-contrast}}
\newcommand{\ControlGroup}{Control}
\newcommand{\Rpref}{\phi_T}
\providecommand{\Description}[1]{}

\title{\textbf{Assessing Policy Updates: Toward Trust-Preserving Intelligent User Interfaces}}

\author{
  Matan Solomon$^{1}$ \quad
  Ofra Amir$^{1}$ \quad
  Omer Ben-Porat$^{1}$\\[4pt]
  $^{1}$Technion –– Israel Institute of Technology, Haifa, Israel\\[2pt]
  \texttt{ma.solomon@campus.technion.ac.il} \quad
  \texttt{oamir@technion.ac.il} \quad
  \texttt{omerbp@technion.ac.il}
}

\date{} 

\begin{document}

\maketitle

\begin{abstract}
Reinforcement learning agents are often updated with human feedback, yet such updates can be unreliable: reward misspecification, preference conflicts, or limited data may leave policies unchanged or even worse. Because policies are difficult to interpret directly, users face the challenge of deciding whether an update has truly helped. We propose that assessing model updates---not just a single model---is a critical design challenge for intelligent user interfaces.
In a controlled study, participants provided feedback to an agent in a gridworld and then compared its original and updated policies. We evaluated four strategies for communicating updates: no demonstration, same-context, random-context, and salient-contrast demonstrations designed to highlight informative differences. Salient-contrast demonstrations significantly improved participants' ability to detect when updates helped or harmed performance, mitigating participants' bias towards assuming that feedback is always beneficial, and supported better trust calibration across contexts.
\end{abstract}

\textbf{Keywords:} Explainability, trust calibration, intelligent user interfaces, human--AI interaction

\section{Introduction}

As reinforcement learning (RL) and machine learning systems increasingly incorporate human feedback, users play an active role in shaping agent behavior \cite{christiano2017deep,macglashan2017interactive}. This human-in-the-loop paradigm is attractive because it allows agents to adapt to individual preferences without requiring a perfectly specified reward function. Yet not all updates from feedback are beneficial: reward misspecification, preference conflicts, and limited training signals can cause agents to adopt behaviors that are unchanged or even worse than before \cite{booth2021rewarddesign,amodei2016concrete}. For effective collaboration, users must therefore be able to assess whether their feedback has actually helped. This challenge is compounded by well-documented tendencies for people to overtrust automated systems or assume that updates are improvements by default \cite{mosier1996automation,parasuraman1997humans}. Without support, such biases risk miscalibrated trust and ineffective delegation.

This assessment step is challenging for two additional reasons. First, designing reward functions in RL is notoriously difficult, making feedback-based updates unreliable and sometimes counterproductive \cite{booth2021rewarddesign}. Second, even when policies are updated, general policies are often opaque to humans: it is hard to predict how a policy will behave across varied contexts simply by inspecting trajectories \cite{amitai2020policies}. As a result, users may default to assuming improvement, which undermines calibrated trust and effective delegation.

While prior work in explainable AI has largely focused on making a single model more interpretable---through feature attributions, saliency maps, or example-based explanations \cite{ribeiro2016lime,shrikumar2017saliency,koh2017influence}---interactive settings introduce a different challenge: users need to understand \emph{how a policy has changed} as a result of their feedback. Recent work on contrastive and comparative explanations \cite{miller2019contrastive,dodge2019contrastive}, as well as approaches that highlight disagreements between policies of different agents \cite{amitai2022disagreements}, point in this direction. Similarly, research on explaining model updates shows that trust depends on how changes are communicated \cite{zhang2020updates,bansal2021updates}. However, these approaches have not been examined in the context of human-in-the-loop RL, where users must repeatedly evaluate policy improvement after intervention.

In this paper, we study how different forms of demonstrations affect users' ability to assess policy updates. In a controlled user study, participants provided feedback to an agent in a grid-world environment, after which they compared the original and updated policies. We evaluated four strategies for communicating updates: (1) no demonstration (\emph{control}), (2) demonstrations in the same context, (3) demonstrations in a random context, and (4) demonstrations in a \emph{salient-contrast} context chosen to highlight the most informative differences between policies. Results show that salient-contrast demonstrations significantly improved participants' ability to detect whether updates helped, especially when policies deteriorated. They also supported better calibration of trust and improved generalization across contexts.

\paragraph{Contributions.} This paper makes three contributions:
\begin{enumerate}
    \item We frame and empirically study the challenge of helping users evaluate whether their feedback meaningfully improves an agent's policy, extending prior work on explainable AI, model updates, and human-in-the-loop RL.
    \item We introduce and compare four strategies for communicating policy updates through side-by-side demonstrations: control, same context, random context, and salient-contrast.
    \item We present evidence from an online user study showing that salient-contrast demonstrations better support accurate evaluation, counteract users' default assumption that feedback always helps, improve trust calibration, and lead to better generalization of agent policies.
\end{enumerate}
Together, these contributions highlight comparative explanations of \emph{change} as a valuable design principle for interactive AI.

\section{Related Work}\label{sec:related}

Our work connects to four strands of research: human-in-the-loop reinforcement learning, challenges of reward design and policy understanding, explanations in interactive AI, and communicating model updates.

\subsection{Human-in-the-Loop Reinforcement Learning}
Human feedback is increasingly used to guide reinforcement learning (RL) agents, either through preferences, demonstrations, or corrective actions. Approaches such as learning from human preferences \cite{christiano2017deep} and policy-dependent corrections \cite{macglashan2017interactive} show that feedback can be effective for shaping policies. Surveys of interactive RL highlight a growing range of mechanisms for incorporating human input \cite{zhang2019irlf,surveyHIL2020}. 

Personalization research in AI and robotics offers complementary strategies for adapting policies to individuals. These include online fine-tuning \cite{yousefi2025policy_finetuning}, selecting from a library of pre-trained agents \cite{dun2023fedjets,ge2025learning_personalization}, and integrating preferences directly into training~\cite{ivanov2024personalized}. In this work, we adopt the second strategy, leveraging a bank of pre-trained policies for tractability, consistent with approaches that approximate the adaptability of fine-tuning without its computational cost. 

While these literatures focus on how agents can learn effectively from humans, less attention has been paid to the inverse question: how users can evaluate whether their feedback actually improved the policy. Our study addresses this gap.

\subsection{Challenges in Reward Design and Policy Understanding}
Designing reward functions in RL is notoriously difficult. Even small mis-specifications can lead to unintended behaviors, a phenomenon documented as reward hacking and specification gaming \cite{amodei2016concrete,krakovna2020specification}. Booth and Shah \cite{booth2021rewarddesign} argue that human feedback alone is insufficient to guarantee alignment, as feedback may also introduce undesirable behaviors. These challenges motivate the need for mechanisms that help users recognize when updates fail to improve. 

A related difficulty is that policies in RL are often hard for humans to interpret. As Amitai and Amir note, general policies cannot easily be summarized by a few trajectories, making it difficult to predict performance across contexts~\cite{amitai2020policies}. They also propose highlighting disagreements between policies as one way to surface behavioral differences \cite{amitai2022disagreements}. Our work builds on these insights but focuses specifically on helping users judge whether a policy has improved after their own feedback.

\subsection{Explanations in Interactive AI}
Explainable AI (XAI) research has proposed a wide range of techniques for making models more interpretable, including feature attributions \cite{ribeiro2016lime,shrikumar2017saliency}, influence functions \cite{koh2017influence}, and example-based explanations \cite{kim2016examples}. However, most approaches focus on explaining a single model. In contrast, interactive settings often require users to compare models across updates. 

Communicating policies through demonstrations has long been explored in RL and robotics. Amir and Amir introduced automated selection of highlight trajectories---segments that are informative or surprising under an agent's $Q$-values \cite{amir2018highlights}. Building on this, Amitai and Amir extended trajectory selection to cross-policy comparison by choosing demonstrations that maximize disagreement between policies \cite{amitai2022disagreements}. Complementary work investigates contrastive and comparative explanations more broadly, showing that highlighting differences improves human judgment \cite{miller2019contrastive,dodge2019contrastive,van2018contrastive_for_RL,krarup2021contrastive_plans}. Our work contributes to this line by evaluating whether contrastively selected demonstrations of pre- and post-update policies help users judge improvement after feedback.

\subsection{Communicating Model Updates and Calibrating Trust}
Users' trust in AI systems depends not only on model accuracy but also on how changes are communicated across updates. Zhang et al.\ \cite{zhang2020updates} show that users often maintain trust even when updates reduce performance. Bansal et al.\ \cite{bansal2021updates} highlight how explanation quality influences team performance with AI. Pareek et al.\ \cite{pareek2024trust_repair} find that even minimal transparency (such as notifying users that a model updated) can aid trust repair after failures. Wang et al.\ \cite{wang2023watch_for_update} demonstrate that explanation changes across updates affect perceived accuracy and consistency. Recent work on model ``diffs'' and auditing further emphasizes the need for transparency when communicating what has changed between models ~\cite{hemmat2025delta}. 

These findings align with broader work on automation bias \cite{mosier1996automation,parasuraman1997humans}, which shows that users often overtrust systems or assume improvement by default. Our study complements this literature by focusing on interactive RL, where user-provided feedback triggers updates, and where it is critical to support calibrated trust through effective demonstrations.

\section{Problem Definition}\label{sec:problem}

When people provide feedback to an adaptive agent, the system updates its policy. A central challenge is that not all updates are beneficial: some improve alignment with user preferences, others leave behavior unchanged, and still others introduce new failures. For effective collaboration, users must be able to \emph{evaluate whether an update has helped}---deciding when to trust the agent and when to intervene further. We refer to this challenge as the problem of \emph{assessing policy updates}.

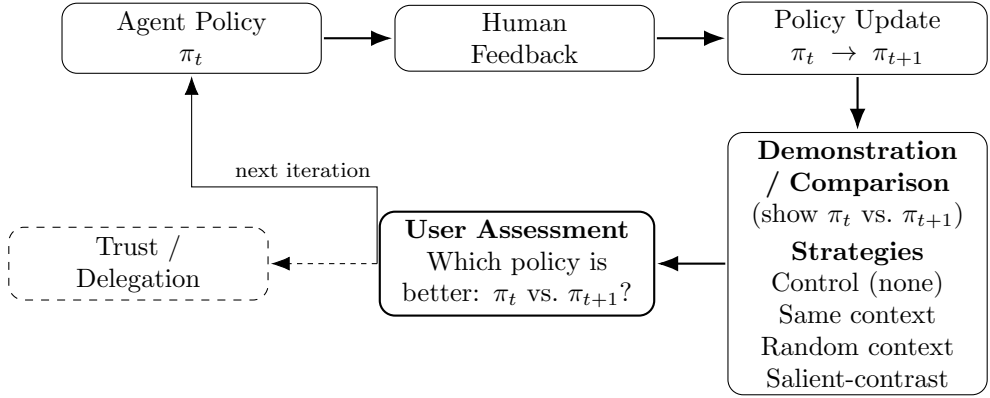
\begin{figure}[t]
\centering
\begin{tikzpicture}[
  font=\small,
  node distance=7mm and 9mm,
  >=Latex,
  box/.style={
    draw, rounded corners=2mm, align=center,
    inner sep=3pt, outer sep=1pt,
    text width=32mm, minimum height=8mm, fill=white
  },
  emph/.style={
    draw, rounded corners=2mm, align=center,
    inner sep=3pt, outer sep=1pt,
    text width=34mm, minimum height=9mm, thick, fill=white
  },
  conceptual/.style={box, dashed},
  flow/.style={-{Latex[length=2.8mm]}, thick},
  thinflow/.style={-{Latex[length=2.5mm]}},
  dashedflow/.style={dash pattern=on 2pt off 2pt, -{Latex[length=2.5mm]}}
]

\node[box] (pit) {Agent Policy\\$\pi_t$};
\node[box, right=of pit] (feedback) {Human\\Feedback};
\node[box, right=of feedback] (update) {Policy Update\\$\pi_t \rightarrow \pi_{t+1}$};

\node[box, below=of update] (demo) {%
\textbf{Demonstration / Comparison}\\
(show $\pi_t$ vs.\ $\pi_{t+1}$)\\[2pt]
\textbf{Strategies}\\
Control (none)\\
Same context\\
Random context\\
Salient-contrast
};

\node[emph, left=of demo] (assess) {\textbf{User Assessment}\\Which policy is better: $\pi_t$ vs.\ $\pi_{t+1}$?};

\node[conceptual, left=of assess, xshift=-5mm] (delegate) {Trust /\\Delegation};

\draw[flow] (pit) -- (feedback);
\draw[flow] (feedback) -- (update);
\draw[flow] (update) -- (demo);
\draw[flow] (demo) -- (assess);

\draw[dashedflow] (assess) -- (delegate);

\draw[thinflow]
  (assess.west)
  |- node[pos=.7, above, sloped, font=\scriptsize] {next iteration}
     ($(delegate.north)+(7mm,5mm)$)
  -| (pit.south);

\end{tikzpicture}
\caption{The feedback–assessment loop in interactive AI. A user observes an agent following policy $\pi_t$, provides feedback, and the agent updates to $\pi_{t+1}$. The system then presents a comparison of $\pi_t$ and $\pi_{t+1}$ using one of four strategies (control, same context, random context, or salient-contrast). The user assesses whether $V^H(\pi_{t+1})>V^H(\pi_t)$ and decides whether to adopt the new policy, informing (conceptual) trust/delegation and further interaction. \emph{Note: Trust/delegation was measured only after multiple rounds, not after each update.}}
\label{fig:feedback-loop}
\end{figure}

Figure~\ref{fig:feedback-loop} illustrates this feedback--assessment loop. A user observes an agent's behavior, provides feedback, and the agent updates its policy. The system then presents a demonstration comparing the old and new policies. The user must assess whether the update improved the policy or whether they wish to revert to the previous policy. This process can potentially be iterated several times. Finally, the user decides whether to delegate to the agent or act on their own.

Formally, we consider an agent acting in an episodic Markov Decision Process (MDP) 
$\langle \mathcal{S}, \mathcal{A}, P, R^H, \gamma \rangle$, where:
\begin{itemize}
  \item $\mathcal{S}$ is the state space,
  \item $\mathcal{A}$ is the action space,
  \item $P: \mathcal{S} \times \mathcal{A} \rightarrow \Delta(\mathcal{S})$ is the transition function,
  \item $R^H: \mathcal{S} \times \mathcal{A} \rightarrow \mathbb{R}$ is the reward function reflecting a user's preferences, and
  \item $\gamma \in [0,1)$ is the discount factor.
\end{itemize}
Given a policy $\pi$, we define its value under user preferences as \[
V^H(\pi) = 
\mathbb{E}_{(s_t,a_t)\sim (\pi)}\!\left[\sum_{t=0}^{\infty} \gamma^t R^H(s_t,a_t)\right],
\]
where the expectation is taken over state–action sequences generated by following $\pi$ in the environment.

An ideal evaluation mechanism would enable the user to judge correctly whether
\[
V^H(\pi_{t+1}) \;>\; V^H(\pi_t),
\]
i.e., whether the updated policy achieves a higher expected cumulative reward. 
In practice, the system can only convey information through demonstrations or explanations, which vary in how effectively they surface improvements or regressions.

\subsection*{Demonstrations as Comparative Explanations of Change}
A key design choice in supporting assessment is \emph{how to demonstrate} the change between $\pi_t$ and $\pi_{t+1}$. We propose \textbf{\Groupname}: side-by-side demonstrations of the two policies in a context where their behaviors differ most. By making differences salient, this approach aims to provide the clearest comparative evidence of improvement or deterioration, thereby best supporting users in assessing updates.

To evaluate the effectiveness of this approach, we compare it against three baseline strategies for selecting demonstrations:

\begin{itemize}
  \item \textbf{Control (no demonstration).} Notification only, providing no evidential support for change.
  \item \textbf{Same context.} Show $\pi_t$ and $\pi_{t+1}$ on the \emph{feedback} board, maximizing local transparency of how feedback affected behavior.
  \item \textbf{Random context.} Show $\pi_t$ and $\pi_{t+1}$ on a board sampled independently of feedback, offering a new perspective but without emphasizing informative differences.
\end{itemize}

These baselines allow us to test whether highlighting salient differences provides unique benefits for users, or whether simpler demonstration strategies are sufficient.

\section{Research Questions and Hypotheses}

Based on the problem definition and prior work, we focus on how different demonstration strategies affect users' ability to assess policy updates. Specifically, we explore two research questions:

\begin{itemize}
  \item[\textbf{RQ1:}] How do different demonstration strategies affect users' ability to determine whether an update improved the agent's policy, both in the \emph{same training context} and across \emph{novel contexts}?
  \item[\textbf{RQ2:}] How do these strategies influence users' broader judgments, including trust calibration and beliefs about the agent's general capabilities?
\end{itemize}

From these questions we derive three hypotheses:

\paragraph{H1: Update Evaluation.} 
Demonstrations that highlight salient differences between policies (\emph{salient-contrast}) will, overall, help participants more accurately judge whether an update improved performance, compared to same or random demonstrations. 
This hypothesis builds on work showing that people reason more accurately when explanations highlight differences. Contrastive explanations---which explain why an outcome occurred rather than an alternative---have been shown to improve users' understanding and fairness judgments \cite{miller2019contrastive,dodge2019contrastive}. In sequential decision-making, highlighting behavioral differences has been argued to support interpretability \cite{juozapaitis2019explainable,madumal2020explainable}. Closer to our setting, Amitai and Amir \cite{amitai2022disagreements} proposed highlighting disagreements between policies of different agents as a way to support user understanding. We extend these insights to the problem of assessing whether a policy has improved after feedback.

\paragraph{H2: Deterioration Sensitivity.} 
The benefit of salient-contrast will be strongest when the updated policy performs worse, countering the human tendency to assume that feedback always helps. 
This hypothesis builds on evidence that people often assume systems improve by default. Prior work on automation bias shows that users tend to overtrust automated systems and accept outputs even when they are erroneous \cite{mosier1996automation,parasuraman1997humans}. Similarly, studies of model updates find that users frequently maintain or even increase trust across updates, regardless of whether performance actually improves \cite{zhang2020updates,bansal2021updates}. This tendency makes it especially important to highlight cases where performance has deteriorated. Research on explanations suggests that contrastive or difference-focused presentations are particularly effective at surfacing such failures: for example, contrastive explanations have been shown to help people detect unfair or problematic outcomes that might otherwise be overlooked \cite{dodge2019contrastive}. In reinforcement learning, misspecified rewards and feedback can worsen policies \cite{booth2021rewarddesign,amodei2016concrete}, yet users may not recognize these regressions without explicit support. We therefore expect salient-contrast demonstrations to be most beneficial when policies deteriorate.

\paragraph{H3: Capability Assessment.} 
Beyond judging specific updates, salient-contrast demonstrations will help participants form more accurate beliefs about the updated agent's general capabilities, leading to better trust calibration and more effective delegation. 
This hypothesis is motivated by research on trust calibration and the difficulty of understanding general policies. Users must often decide not only whether a specific update helped but also whether to rely on the agent in future contexts. Prior work shows that explanations can help people decide when to trust an AI system \cite{bansal2021updates}, yet users struggle to recalibrate trust across model updates unless differences are made salient \cite{zhang2020updates}. Moreover, policies in reinforcement learning are inherently difficult to understand in general environments \cite{amitai2020policies}. Highlighting disagreements between policies, as suggested by Amitai and Amir \cite{amitai2022disagreements}, provides one approach for surfacing general behavioral differences. We therefore expect that salient-contrast demonstrations will better support participants in projecting an agent's capabilities beyond the training context, improving trust calibration and delegation.

\section{Empirical Methodology}\label{sec:method}
We ran a user study to test the hypotheses regarding the different policy update demonstration strategies. In the study, participants provided feedback to an agent acting in a grid-world, and then reviewed and assessed the updated policies. The study has been approved by our Institutional Review Board.

\subsection{Empirical Domain}
\begin{figure}[t]
  \centering
  \includegraphics[width=0.42\linewidth]{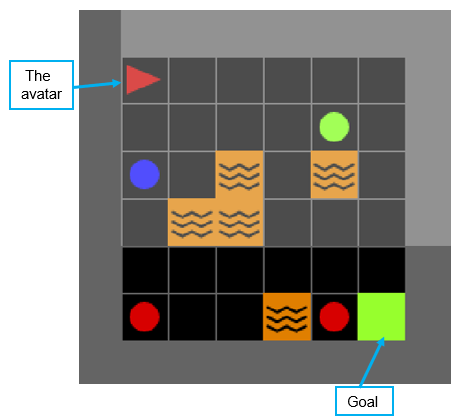}
  \caption{Example of an initial state of the environment used in the study.}
  \label{fig:initial_state}
\end{figure}

We investigate users' evaluations of policy updates in a finite-horizon gridworld based on MiniGrid~\cite{chevalier2024minigrid}. The environment consists of a discrete grid containing collectible colored balls (Blue, Green, Red), lava hazard tiles, and a terminal goal tile. Figure~\ref{fig:initial_state} shows a screen of one of the initial states we used in the experiment. At each timestep, the agent selects one of four discrete actions,
\[
\mathcal{A}=\{\text{Forward},\ \text{TurnRight},\ \text{TurnLeft},\ \text{Pickup}\},
\]
and an episode terminates upon reaching the goal or after $T_{\max}=70$ steps.

Task performance is scored by a linear reward model over five features. Let $\phi\in\mathbb{R}^{5}$ encode (i) rewards for collecting each of the three balls, (ii) a penalty for entering lava, and (iii) a per-step cost. The evaluation reward shown to participants is
\[
\phi_T = (4,\ 2,\ -2,\ -3,\ -0.1),
\]
corresponding respectively to (Blue, Green, Red, Lava, Step Cost).

To obtain policy variants that reflect different inductive biases, we pre-trained $k=6$ policies with PPO~\cite{schulman2017ppo}, each optimized to convergence under a distinct preference vector $\phi^{(i)}\in\mathbb{R}^{5}$. These vectors vary the relative weighting of hazards and collectibles (e.g., strong vs.\ weak lava aversion and heterogeneous ball preferences), yielding a bank of agents with systematically different behaviors evaluated under the common $\Rpref$.

\subsection{Experimental Conditions}
After each update, participants saw a demonstration comparing the old and updated policies, with the board (i.e., grid-world specification) determined by a between-subject condition:
\begin{itemize}
  \item \textbf{\ControlGroup:} Text notice that the agent was updated; the system advanced with the new policy (no A/B choice).
  \item \textbf{Same:} Side-by-side demonstration on the \emph{same} board on which feedback was given.
  \item \textbf{\Groupname:} Demonstration on a board that maximizes the score difference between the two policies under $\Rpref$.
  \item \textbf{Random:} Side-by-side demonstration on a board drawn uniformly at random from a pre-generated pool of 18 evaluation boards ($\mathcal{S}_{18}$), curated to span diverse initial layouts and elicit distinct policy behaviors; draws were independent of the feedback board.
\end{itemize}

\subsection{Procedure}
The study comprised three stages executed across five update cycles.

\paragraph{Stage 1 — Familiarization.}
Participants first completed a brief tutorial covering the environment dynamics, action set, and evaluation rewards, followed by a short comprehension check (quiz) to verify understanding. They then played three practice episodes under manual control to further familiarize themselves with the environment.

\begin{figure}[H]
  \centering
  \includegraphics[width=0.62\linewidth]{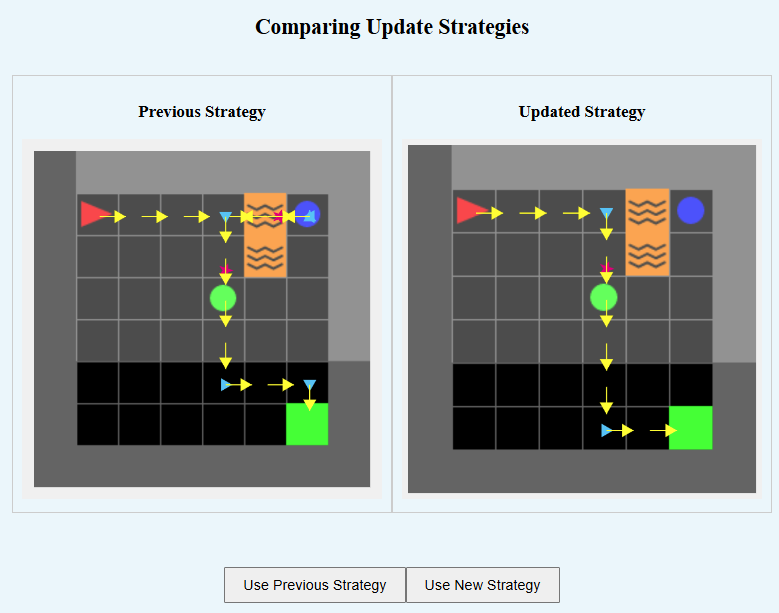}
  \caption{Side-by-side demonstration of the old policy and the updated policy during Stage~2.}
  \label{fig:compare_strategy}
\end{figure}

\paragraph{Stage 2 — Corrective Feedback and Update.}
In each of five rounds (see Algorithm~\ref{alg:policy_update_cycle}):
\begin{enumerate}
  \item \textit{Observe.} The participant watched the current agent $\pi_t$ act on a fixed board.
  \item \textit{Correct.} The participant reviewed the most recent episode trajectory. Whenever the agent's action $\pi_t(s_t)=a_b$ differed from the participant's preferred action $a_p$, the participant supplied $a_p$. We logged tuples $(s_t,a_b,a_p)$.
  \item \textit{Select Update.} From a neighborhood of pre-trained policies, we selected the candidate $\pi_{\text{new}}$ that maximized agreement with the cumulative corrections.
  \item \textit{Demonstrate.} Before comparison, participants were explicitly told that policy updates can improve or degrade performance. We then presented a side-by-side demonstration of the incumbent policy $\pi_t$ and the candidate update $\pi_{\text{new}}$ on a board selected according to the assigned condition (see Fig.~\ref{fig:compare_strategy}).
  \item \textit{Choose.} The participant adopted or rejected the update (the \textit{\ControlGroup} condition advanced automatically with the update).
\end{enumerate}

\begin{algorithm}[t!]
\caption{Policy selection with corrective feedback (five sessions)}
\label{alg:policy_update_cycle}
\begin{algorithmic}[1]
\STATE \textbf{Input:} Initial policy $\pi_0$; pre-trained bank $\{\pi^{(j)}\}_{j=1}^m$; neighborhood operator $\mathcal{N}(\cdot)$
\STATE \textbf{Initialize:} $\pi \leftarrow \pi_0$
\FOR{$\text{round} = 1$ \TO $5$}
  \STATE Run $\pi$; collect feedback $\mathcal{D} \gets \{(s_t, a_b, a_p)\}$
  \STATE $C \gets \mathcal{N}(\pi) \subseteq \{\pi^{(j)}\}_{j=1}^m$
  \STATE For each $\tilde{\pi}\in C$, compute
  \[
    \mathrm{Agree}(\tilde{\pi}) \;=\; \sum_{(s_t,a_b,a_p)\in \mathcal{D}} \mathbf{1}\!\left[\arg\max_a \tilde{\pi}(a\mid s_t)=a_p\right].
  \]
  \STATE $\pi_{\text{new}} \leftarrow \arg\max_{\tilde{\pi}\in C}\ \mathrm{Agree}(\tilde{\pi})$
  \STATE Demonstrate $\pi$ vs.\ $\pi_{\text{new}}$ according to the assigned condition
  \STATE \textbf{if} participant accepts \textbf{then} $\pi \leftarrow \pi_{\text{new}}$
\ENDFOR
\end{algorithmic}
\end{algorithm}

\paragraph{Stage 3 — Trust Evaluation.} After completing the five feedback rounds, participants faced an incentivized delegation decision: their payment bonus was proportional to the game score in the next round, and they chose whether the trained agent ($\pi_{\text{final}}$) or they themselves should play. We also collected the participants' evaluation of the trained agent on a 7-point Likert scale. Finally, participants—excluding those in \ControlGroup, who did not receive a demonstration—completed an explanation-satisfaction questionnaire adapted from Hoffman et al.~\cite{hoffman2018metrics}.

\subsection{Evaluation Metrics and Analyses}
To test hypotheses H1 (updated evaluation), H2 (deterioration sensitivity),  we use the following measures:  
  \begin{itemize}
    \item \emph{Correct Choice — Generalized}: Indicator of whether the participant selected the better policy when performance is averaged over the fixed suite of $18$ initial boards $\mathcal{S}_{18}$ under $\Rpref$.
    \item \emph{Correct Choice — Local}: Indicator of whether the participant selected the better policy only on the \emph{feedback board} used in that session, scored under $\Rpref$. 
\end{itemize}

To test H3 (capability assessment), we assess the following measures:
  \begin{itemize}
    \item \emph{Trust the Agent}: Final delegation decision in Stage~3 (agent vs.\ self).
    \item \emph{Agent Evaluation}: Self-report (1--7) comparing the agent's skill to the participant (1= ``much worse'', 4= ``same'', 7=``much better'').
  \end{itemize}
  We also compute the quality of the final agent reached by participants (i.e., expected return of $\pi_{\text{final}}$ under $\Rpref$ over $\mathcal{S}_{18}$).

Beyond testing the specified hypotheses, we also measure and analyze the following:
\begin{itemize}
    \item \textbf{Explanation satisfaction.} We adapt the explanation satisfaction scale from Hoffman at el.~\cite{hoffman2018metrics} (4 items- understandable, overwhelming, feedback-contribution, helpful).
\end{itemize}

  \textbf{Statistical plan.} Unless noted otherwise, between-group comparisons use Mann–Whitney $U$ tests; we report significant findings in the main text and provide full descriptive plots and confirmatory tests in the Results section. We used a one-sided test in comparisons that followed a directional hypothesis. All data is available (anonymized) on \url{https://osf.io/jcxyk/?view_only=69fed692771a493f92f6559bc0f640b1}.

\noindent\textbf{Participants.} We recruited $N=308$ Prolific participants (median completion time $\approx 25$ minutes; £7/hour + up to £1 bonus). We excluded (i) participants who provided no relevant corrective feedback and (ii) those who failed the attention check, yielding $N=280$ ($M_{\text{age}}=41.1$, $SD=12.12$). Randomization yielded the following allocation across conditions: \ControlGroup~ $n=69$, Same $n=62$, \Groupname~ $n=73$, Random $n=76$.

\section{Results}
We begin by describing the results with respect to the stated hypotheses, and then describe explanation satisfaction results.

\subsection{Update evaluation (H1)}
\begin{figure}[t]
  \centering
  \begin{subfigure}[t]{0.48\linewidth}
    \centering
    \includegraphics[width=\linewidth]{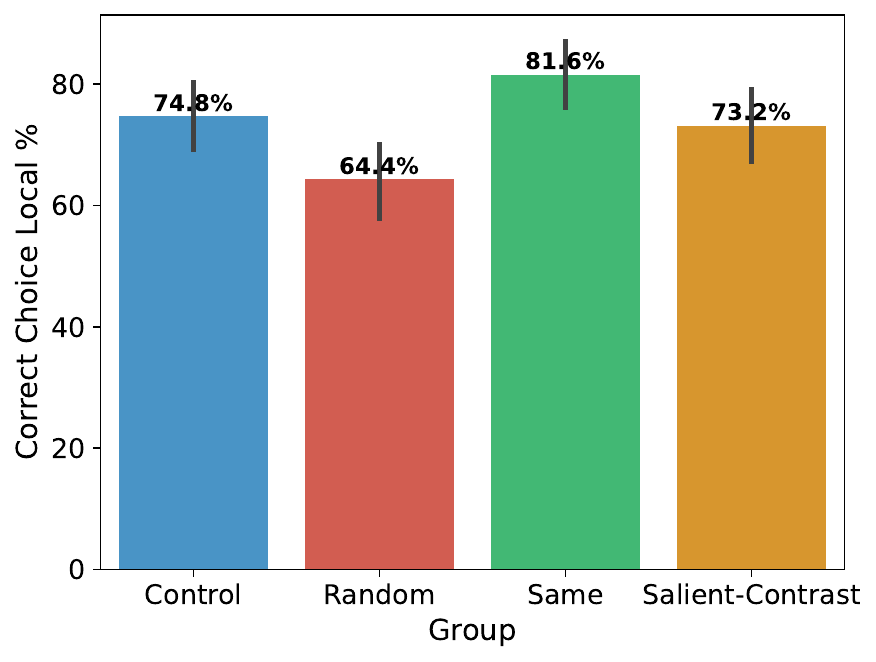}
    \caption{Local (feedback board).}
    \label{fig:correct-choice-feedback}
  \end{subfigure}\hfill
  \begin{subfigure}[t]{0.48\linewidth}
    \centering
    \includegraphics[width=\linewidth]{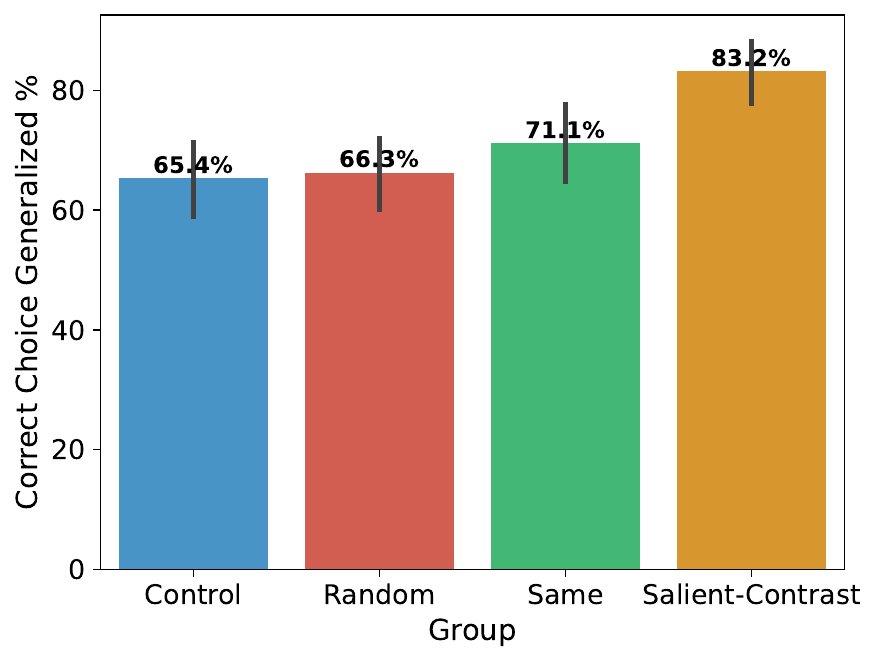}
    \caption{Generalized (across boards).}
    \label{fig:correct-choice-mean}
  \end{subfigure}

  \caption{\textbf{Correct agent choice across conditions.}
  Two bar charts compare the four experimental groups.
  (a)~Local correctness is computed on the feedback board, where the \emph{Same} condition achieves the highest accuracy. 
  (b)~Generalized correctness is averaged across a fixed 18-board set under $\Rpref$, where \Groupname\ attains the highest scores. 
  These results illustrate that same-context feedback supports local assessment, while salient-contrast demonstrations promote generalization. 
  Error bars denote 95\% confidence intervals.}
  \label{fig:correct-choice-sidebyside}
\end{figure}

We found that participants in the \Groupname\ group were more accurate in assessing whether updates led to a better policy, as shown in 
Figure~\ref{fig:correct-choice-mean}. \Groupname\ achieved the highest correct-choice rate, significantly exceeding \textit{Same} (Mann–Whitney $U=24771$, $p=0.003$), \textit{Random} ($U=23858$, $p<0.0001$), and \textit{\ControlGroup} ($U=21159$, $p<0.0001$). These results align with \textbf{H1}: contrastive demonstrations that highlight salient differences support generalization beyond the specific feedback instance.
As a descriptive reference point, in the \textit{\ControlGroup}—where updates were automatically adopted—performance improved in 74\% of comparisons on the feedback board and in 65\% when averaged across boards, indicating that updates often (but not always) helped.

In contrast, participants in the \textit{Same} condition outperformed the others when assessing the update locally, i.e., whether the policy improved for the board in which the feedback was provided: vs.\ \Groupname\ (Mann–Whitney $U=20250$, $p=0.04$), vs.\ \textit{Random} ($U=21712$, $p<0.0001$), and vs.\ \textit{\ControlGroup} ($U=21916$, $p=0.088$), see Figure~\ref{fig:correct-choice-feedback}. This pattern is expected: repeating the exact context in which feedback was provided affords a local advantage. However, it was less useful in helping participants assess whether the policy improved across different settings.

\subsection{Deterioration sensitivity (H2)}
To test \textbf{H2}, we stratified trials by update direction (\emph{Positive} if $\pi_{t+1}$ outperformed $\pi_t$ under $\Rpref$, \emph{Negative} otherwise) and compared \textit{Same} to \Groupname.

\begin{figure}[H]
  \centering
  \begin{subfigure}[t]{0.48\linewidth}
    \centering
    \includegraphics[width=\linewidth]{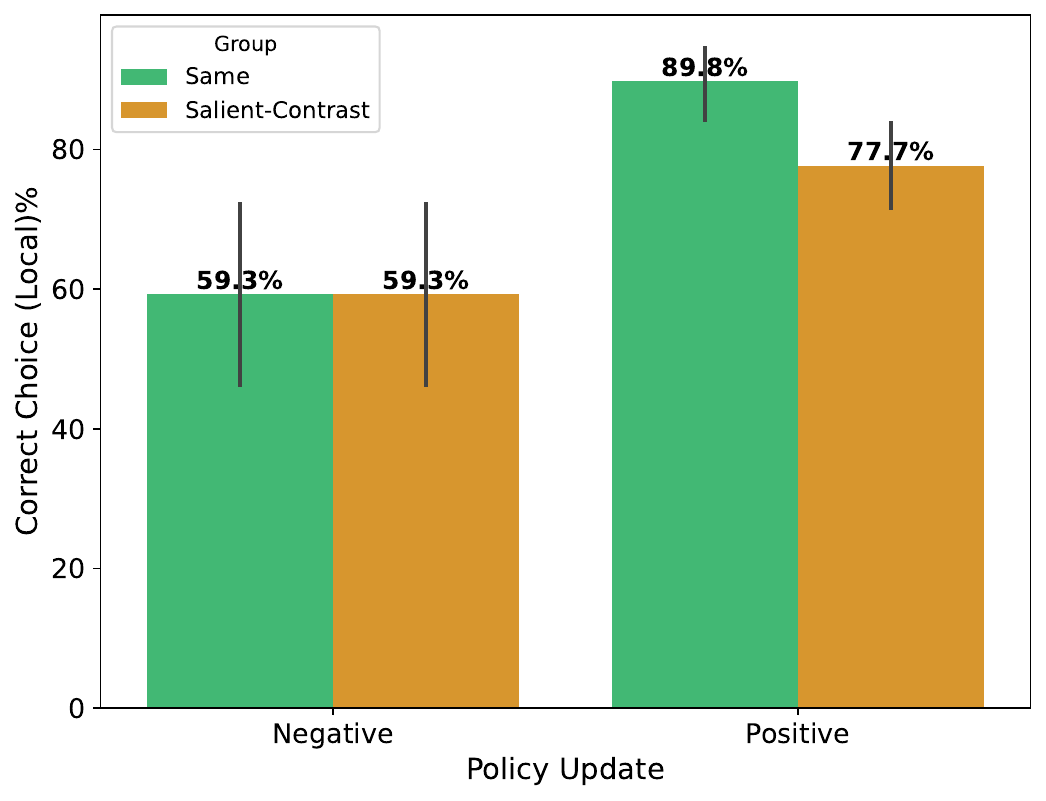}
    \caption{Local (feedback board).}
    \label{fig:choice-local-split}
  \end{subfigure}\hfill
  \begin{subfigure}[t]{0.48\linewidth}
    \centering
    \includegraphics[width=\linewidth]{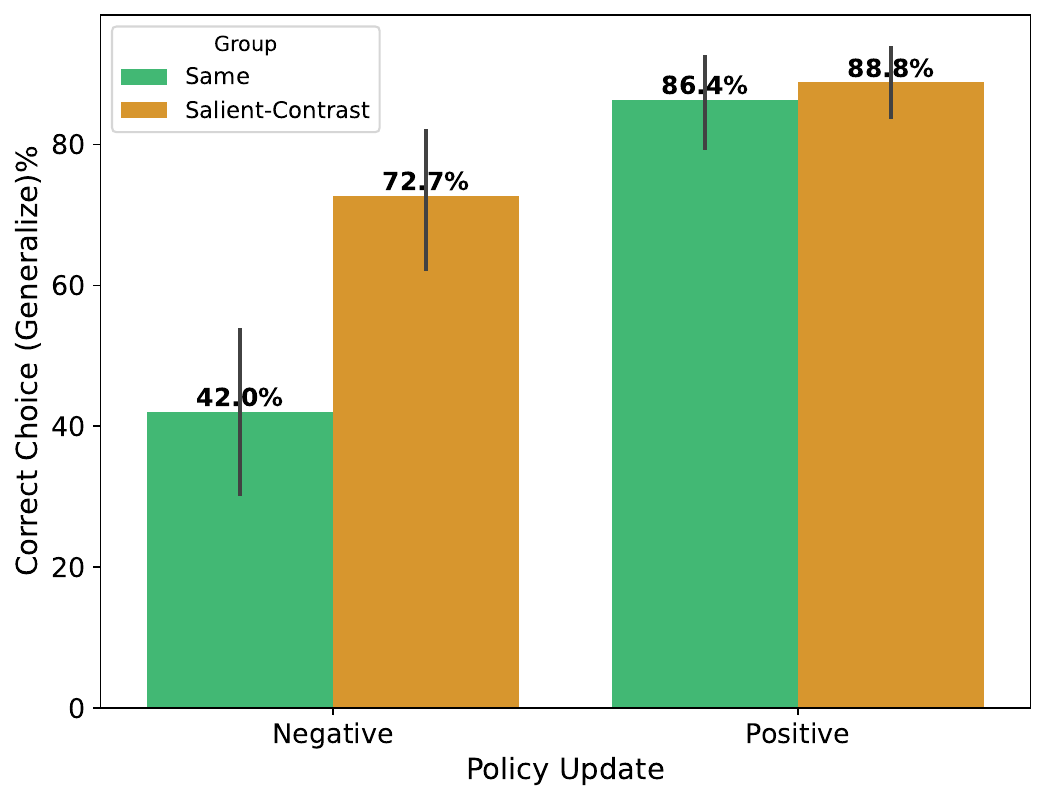}
    \caption{Generalized (across boards).}
    \label{fig:choice-mean-split}
  \end{subfigure}

  \caption{\textbf{Correct agent choice by update direction (\textit{Same} vs.\ \Groupname).}
  Each panel shows the proportion of correct update evaluations split by whether the update improved or worsened performance. 
  (a)~On positive updates, the \emph{Same} group achieves higher local correctness. 
  (b)~On negative updates, \Groupname\ shows substantially higher generalized correctness, indicating stronger ability to detect and reject harmful updates. 
  Error bars denote 95\% confidence intervals.}
  \label{fig:correct-choice-separated-sidebyside}
\end{figure}

For \emph{Correct Choice—Local} (Fig.~\ref{fig:choice-local-split}), \textit{Same} exceeded \Groupname\ on \emph{Positive} updates (89.8\% vs.\ 77.7\%; Mann–Whitney $U=10726$, $p=0.004$), consistent with its local-context advantage; no reliable difference emerged on \emph{Negative} updates.

By contrast, for \emph{Correct Choice—Generalize} (Fig.~\ref{fig:choice-mean-split}), results reversed on \emph{Negative} updates: \Groupname\ substantially outperformed \textit{Same} (72.7\% vs.\ 42.0\%; $U=3472$, $p=0.0002$), with only a minor advantage for \Groupname\ on \emph{Positive} updates. These findings strongly support \textbf{H2}: salient-contrast demonstrations are especially effective at helping participants \emph{detect and reject} harmful updates, while same-context demonstrations primarily surface \emph{local} improvements.

\subsection{Capability assessment (H3)}
\begin{figure}[t]
  \centering
  \begin{subfigure}[t]{0.48\linewidth}
    \centering
    \includegraphics[width=\linewidth]{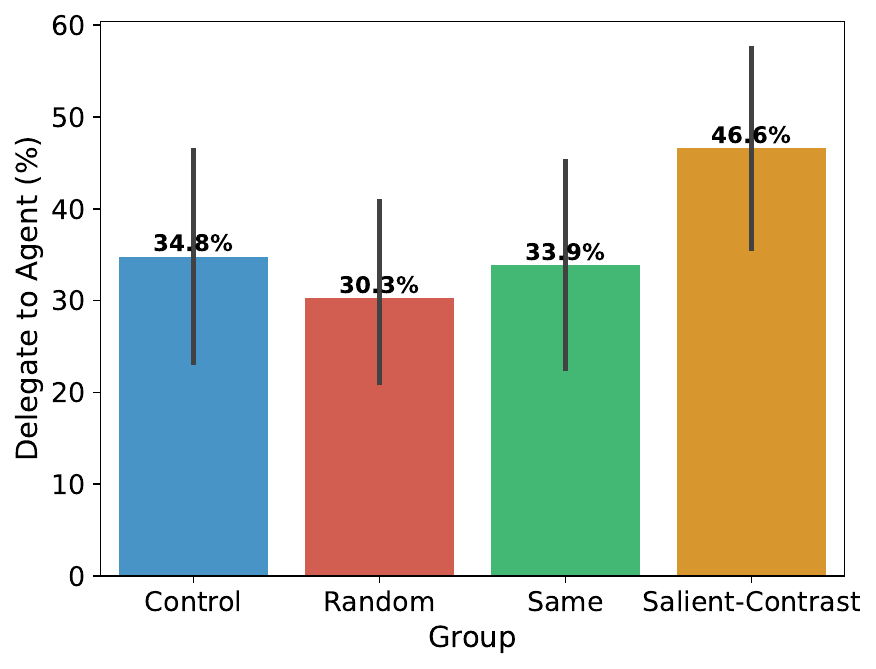}
    \caption{Delegation (\emph{Trust the Agent}) across groups.}
    \label{fig:trust}
  \end{subfigure}\hfill
  \begin{subfigure}[t]{0.48\linewidth}
    \centering
    \includegraphics[width=\linewidth]{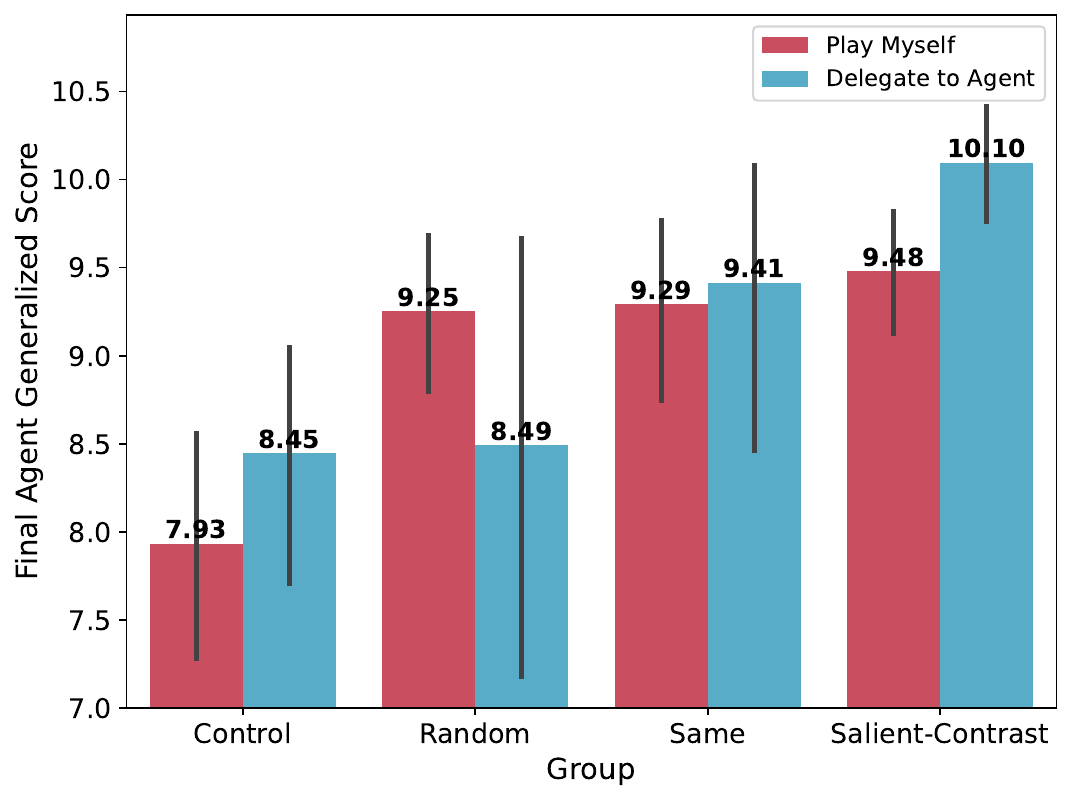}
    \caption{Final-agent generalized score by delegation and condition.}
    \label{fig:final_agent_scores}
  \end{subfigure}

  \caption{\textbf{Capability assessment and trust calibration (H3).}
  (a)~Delegation rates are highest in the \Groupname\ condition, indicating greater trust in the trained agent. 
  (b)~Final-agent generalized scores, measured under $\Rpref$, are also highest in \Groupname, with a significant delegate\,>\,self gap only in this condition.}
  \label{fig:trust-and-performance}
\end{figure}

\textbf{Objective agent evaluation.} We quantified capability using each participant's \emph{final-agent generalized score}. \Groupname\ yielded the highest scores overall ($M=9.77$), significantly exceeding \textit{Random} ($M=9.02$, $U=2249$, $p=0.037$) and \textit{\ControlGroup} ($M=8.11$, $U=1278$, $p<0.0001$). Moreover, within \Groupname, participants who delegated achieved higher final-agent scores than those who did not ($U=449$, $p=0.006$). This divergence was not observed in the other conditions (see Fig. ~\ref{fig:final_agent_scores}), suggesting that salient-contrast demonstrations supported both \emph{selection} of better agents and \emph{appropriate reliance} on them.

\textbf{Subjective agent evaluation.} Participants' self-assessments placed the trained agent below their own ability on average (means $<4$ on a 1–7 scale). This pattern is not surprising as we deliberately designed the task such that participants could relatively easily detect suboptimal agent actions and provide effective corrective feedback. Consistent with the objective results, \Groupname\ received higher capability ratings than \textit{Same} ($M=3.64$ vs.\ $M=3.13$, One-sided Mann–Whitney $U=2693$, $p=0.026$) and \textit{\ControlGroup} ($M=3.09$, $U=1995$, $p=0.014$).

\textbf{Delegation outcomes.}
Consistent with \textbf{H3} (Capability Assessment), Figure~\ref{fig:trust} shows that final-stage delegation to the agent was highest in \Groupname. One-sided Mann–Whitney tests (directional per H3) indicate that \Groupname\ exceeded \textit{Random} ($U=2321$, $p=0.02$) and was marginally higher than \textit{Same} ($U=2550$, $p=0.068$) and \textit{\ControlGroup} ($U=2221$, $p=0.078$). Overall, the pattern shows directional evidence consistent with H3: salient-contrast demonstrations help participants \emph{distinguish} higher-quality agents and \emph{delegate} accordingly, supporting more accurate capability assessment and better-calibrated trust.

\subsection{Explanation satisfaction}
Participants, excluding \ControlGroup, completed a multi-item explanation-satisfaction (ES) questionnaire; the scale demonstrated good internal consistency (Cronbach's $\alpha=0.825$, $95\%$ CI $[0.783,\,0.861]$).  Composite \textit{Explanation Satisfaction} means were comparable across conditions: \textit{Same} $M=4.98$, \textit{\Groupname}\ $M=4.90$, and \textit{Random} $M=4.61$, with no statistically significant differences among these means. On the specific ``feedback contribution” item (``The explanations helped me see how my feedback changed the agent's behavior''), two-sided Mann–Whitney tests indicated significant differences when comparing \textit{Same} ($M=5.06$) to \Groupname\ ($M=4.55$, $U=1810$, $p=0.042$) and to \textit{Random} ($M=4.47$, $U=1890$, $p=0.042$).
 This is unsurprising as viewing the exact board on which feedback was provided makes the feedback–behavior linkage especially transparent. However, with respect to \emph{objective} decision quality, both the evaluation of policy improvement (Fig.~\ref{fig:correct-choice-mean}) and the final agents' generalization scores, \Groupname\ surpass \textit{Same}.

\section{Discussion}
\paragraph{Summary and Conclusions}
Our results provide evidence for the value of comparative demonstrations in human-in-the-loop reinforcement learning. Consistent with H1, participants in the \emph{salient-contrast} condition were more accurate in judging whether an update improved performance than those in the same or random conditions. In line with H2, this advantage was strongest when the updated policy actually deteriorated: in such cases, participants in other conditions often assumed improvement by default, whereas salient-contrast displays helped them recognize decline. H3 was partially supported. Salient-contrast demonstrations led to better calibration of trust and higher generalization performance of the final agent. 

These findings extend prior work on explanation satisfaction and model updates. While participants rated the \emph{same} condition highest for linking feedback to behavior, objective performance was superior in the \emph{salient-contrast} condition. This echoes Hoffman et al.'s observations that subjective explanation satisfaction is not always aligned with quality \cite{hoffman2018metrics}, and reinforces prior findings that explanations can affect perceived trust more than actual calibration \cite{bansal2021updates,zhang2020updates}. Our results also provide a concrete method for counteracting automation bias \cite{mosier1996automation,parasuraman1997humans}. By surfacing differences when policies get worse, salient-contrast demonstrations prevent users from overtrusting agents simply because an update occurred. Conceptually, this extends the literature on contrastive and comparative explanations \cite{miller2019contrastive,dodge2019contrastive,amitai2022disagreements} from justifying individual predictions to communicating change across model states.

\paragraph{Design implications.} For designers of interactive AI systems, our findings highlight three practical lessons. First, replaying feedback in the same context may satisfy users but provides limited support for judging whether an update generalizes. Second, demonstrations that highlight salient differences are most valuable when performance deteriorates, making them essential for preventing overtrust. Third, systems should consider hybrid strategies: combining same-context demonstrations to reassure users about local changes with salient-contrast demonstrations to reveal global improvements or failures. More broadly, reinforcing prior work~\cite{buccinca2020proxy}, explanations should be evaluated not only for satisfaction but for their ability to support calibrated trust and informed delegation.

\paragraph{Limitations.}
Several limitations of our study should be noted. Our grid-world setting with pre-trained policies provided experimental control but lacks the richness of real-world domains. Our approach to selecting salient-contrast demonstrations relied on heuristics that may not generalize across environments. We also measured trust and delegation only after several rounds, rather than continuously, which may understate the dynamics of how assessment influences reliance on the agent. Finally, our participant population was not composed of domain experts, so results may differ in professional or high-stakes contexts.

\paragraph{Future work} There are several avenues for future work extending our findings. One avenue is to design adaptive demonstration-selection methods that respond to user performance or highlight differences most likely to be misunderstood. Another is to test these strategies in domains such as robotics or decision support, where policy generalization is critical. Longitudinal studies could explore how assessment strategies influence trust calibration over extended interaction. A promising direction is to combine same-context and salient-contrast approaches, providing both local transparency about how feedback was used and global evidence about how the policy has changed.  

\section{Conclusion}
Taken together, our results underscore the importance of comparative explanations that make policy change salient. Such explanations help users avoid overtrusting deteriorated agents, support more accurate evaluation, and lead to stronger collaborative outcomes. For the IUI community, this work contributes empirical evidence on how demonstration strategies affect human assessment of policy updates, highlights a trade-off between subjective explanation satisfaction and objective decision quality, and introduces comparative explanation design as a principle for building interactive AI systems that better support trust calibration and human--AI collaboration.

\bibliographystyle{unsrt}
\bibliography{bibliography}

\end{document}